\documentclass[prx,twocolumn,superscriptaddress,longbibliography]{revtex4-1}

\usepackage{graphicx,color}
\usepackage{amsmath,amssymb}
\usepackage{textcomp}
\usepackage{braket,slashed}
\usepackage{hyperref}
\usepackage[english]{babel}

\usepackage{bm}

\usepackage[mathscr]{euscript}
\usepackage{dsfont}

\newcommand{\vectorspace}[1]{\ensuremath{\mathbb{#1}}}

\DeclareMathOperator{\tr}{tr}

\newcommand{\ec}{\ensuremath{\mathrm{e}}}

\newcommand{\rket}[1]{\ensuremath{|#1)}}
\newcommand{\rbra}[1]{\ensuremath{(#1|}}
\newcommand{\rbraket}[1]{\ensuremath{(#1)}}

\newcommand{\hilbert}{\ensuremath{\vectorspace{H}}}

\begin{document}
\title{Gauging quantum states: from global to local symmetries in many-body systems}

\author{Jutho Haegeman}
\affiliation{Department of Physics and Astronomy, University of Ghent, Krijgslaan 281 S9, B-9000 Ghent, Belgium}
\author{Karel Van Acoleyen}
\affiliation{Department of Physics and Astronomy, University of Ghent, Krijgslaan 281 S9, B-9000 Ghent, Belgium}
\author{Norbert Schuch}
\affiliation{JARA Institute for Quantum Information, RWTH Aachen University, D-52056 Aachen, Germany}
\author{J. Ignacio Cirac}
\affiliation{Max-Planck Institut f\"{u}r Quantenoptik, Hans-Kopfermann-Str. 1, D-85748 Garching, Germany}
\author{Frank Verstraete}
\affiliation{Department of Physics and Astronomy, University of Ghent, Krijgslaan 281 S9, B-9000 Ghent, Belgium}
\affiliation{Faculty of Physics, University of Vienna, Boltzmanngasse 5, A-1090 Wien, Austria}

\begin{abstract}
We present an operational procedure to transform global symmetries into local symmetries at the level of individual quantum states, as opposed to typical gauging prescriptions for Hamiltonians or Lagrangians. We then construct a compatible gauging map for operators, which preserves locality and reproduces the minimal coupling scheme for simple operators. By combining this construction with the formalism of projected entangled-pair states (PEPS), we can show that an injective PEPS for the matter fields is gauged into a $\mathsf{G}$-injective PEPS for the combined gauge-matter system, which potentially has topological order. We derive the corresponding parent Hamiltonian, which is a frustration free gauge theory Hamiltonian closely related to the Kogut-Susskind Hamiltonian at zero coupling constant. We can then introduce gauge dynamics at finite values of the coupling constant by applying a local filtering operation. This scheme results in a low-parameter family of gauge invariant states of which we can accurately probe the phase diagram, as we illustrate by studying a $\mathbb{Z}_{2}$ gauge theory with Higgs matter.
\end{abstract}

\maketitle
The fascinating subject of gauge theories is omnipresent throughout many-body physics. The gauge principle, which states that the fundamental interactions of nature originate from gauging global symmetries of the free theory, is one of the cornerstones of the Standard Model, but quantized gauge fields also emerge as effective degrees of freedom in several models for strongly correlated condensed matter, alongside other effective interactions. Historically, the concept of \emph{gauging}, \textit{i.e.} transforming a global symmetry into a local symmetry, is based on a Lagrangian or Hamiltonian description of the system where operators of the original (matter) theory are transformed into gauged operators using the `minimal coupling rule', which is not always unambiguous \cite{2013JHEP...09..063J}. Indeed, while there is a unique way to \emph{ungauge} a theory (by setting the gauge fields and gauge coupling constant equal to zero), the reverse process is not unique as new degrees of freedom are introduced and the Hilbert space is enlarged.

The concept of \emph{gauging} is however not strictly tied to any specific dynamics of the matter fields. This manuscript therefore explores how to gauge global symmetries at the level of individual quantum many body states, independent of any prescribed Hamiltonian or Lagrangian. We thereto consider the Hilbert space $\hilbert^{(\text{m})}$ of a quantum many body system living on the vertices of a graph (the matter) and which has a a global action of a group $\mathsf{G}$ defined. We then introduce other degrees of freedom (the gauge field) by enlarging the Hilbert space to $\hilbert^{(\text{g,m})}$ and define a map $G:\hilbert^{(\text{m})}\to\hilbert^{(\text{g,m})}$ which explicitly transforms every matter state $\ket{\psi}$ which is invariant under the global action of $\mathsf{G}$ to a corresponding gauge-matter state $\ket{\Psi}=G\ket{\psi}$ which is invariant under local symmetry actions of $\mathsf{G}$. Only thereafter do we introduce an associated gauging map $\mathscr{G}$ for operators $O$, such that $G O\ket{\psi} = \mathscr{G}[O] G\ket{\psi}$ and local matter operators are mapped to local gauge-matter operators. This map reproduces the well-known result for simple operators such as hopping interactions or correlation functions but also produces unambiguous, operationally defined results for more complex operators involving \textit{e.g.} plaquette interactions for the matter fields.

To this date, the most accurate description of the strongly coupled, nonperturbative behavior of quantum gauge theories comes from Monte Carlo sampling of the path integral corresponding to Wilson's lattice gauge theory (LGT) formulation \cite{PhysRevD.10.2445}. The Hamiltonian formulation of Wilson's LGT, originally developed by Kogut and Susskind \cite{1975PhRvD..11..395K} (but see also \cite{Horn1981149,Orland1990647,1997NuPhB.492..455C}), has been investigated in the context of approximate wave function ans\"{a}tze \cite{Greensite1979469,Feynman1981479,Karabali1998103,PhysRevD.52.3719} and is receiving a renewed interest in the context of cold atom simulators \cite{PhysRevLett.95.040402,PhysRevLett.110.125304,PhysRevLett.112.201601,2013NatCo...4E2615T} and tensor network approaches \cite{2002PhRvD..66a3002B,2005JHEP...07..022S,PhysRevB.83.115127,2013JHEP...11..158B,2013arXiv1312.6654B,2014arXiv1403.0642S,2014arXiv1404.7439S,2014arXiv1405.4811T,Milsted:aa}. The representation of quantum many body states as tensor networks \cite{2008AdPhy..57..143V,2009JPhA...42X4004C} originates from White's successful density matrix renormalization group \cite{1992PhRvL..69.2863W} and is now well established in the context of one-dimensional quantum chains. However, recent results for the tensor network description of higher dimensional quantum systems \cite{2004cond.mat..7066V} (including fermions \cite{2009PhRvB..80p5129C,2010PhRvA..81a0303C,2010PhRvB..81p5104C,2010PhRvA..81e2338K,2014arXiv1404.5268P}), quantum chemistry models \cite{1999JChPh.110.4127W,chan2011density,2014arXiv1403.0981M} and even quantum field theories \cite{2010PhRvL.104s0405V,2010PhRvL.105y1601H} look equally promising. In addition, the theoretical underpinning of tensor network states in terms of the area law of entanglement entropy \cite{2007JSMTE..08...24H,2010RvMP...82..277E} makes them suitable for theoretical results as profound as the complete classification of gapped quantum phases \cite{PhysRevB.84.165139,PhysRevB.84.235128}. The unifying theme in these studies does indeed correspond to a shift in focus from a Hamiltonian or Lagrangian description of the system towards a description in terms of the universal properties of the quantum (ground) state itself.

Following some early results \cite{2002PhRvD..66a3002B,2005JHEP...07..022S}, the application of tensor network states to systems with gauge symmetry has recently seen a revived interest. Aside from some extremely accurate results for the Schwinger model \cite{2013JHEP...11..158B,2013arXiv1312.6654B,2014arXiv1403.0642S}, there have been first explorations with two-dimensional pure gauge theory \cite{PhysRevB.83.115127} and theoretical formulations of classes tensor network states with explicit gauge invariance \cite{2006PhRvL..96v0601V,2014arXiv1404.7439S,2014arXiv1405.4811T,Milsted:aa}. In this manuscript, we also combine our gauging construction with the formalism of projected entangled-pair states \cite{2004cond.mat..7066V} (PEPS). We prove that an injective PEPS \cite{2007arXiv0707.2260P} with global symmetry $\mathsf{G}$ is gauged into a $\mathsf{G}$-injective PEPS \cite{2010AnPhy.325.2153S} (and refer to the Supplementary Material for a summary of these concepts), which reestablishes the close relation between deconfinement and topological order in the case of discrete groups \cite{2003AnPhy.303....2K,1367-2630-13-2-025009}, or compact groups broken down to discrete subgroups \cite{PhysRevLett.62.1221,Preskill199050,AlexanderBais199263}. We explicitly derive a parent Hamiltonian of the gauged PEPS, which resembles the Kogut-Susskind hamiltonian at zero coupling.  We discuss a well known approach for introducing gauge dynamics at nonzero coupling constant and apply this prescription to obtain a low-parameter family of gauge invariant tensor network states that allows for accurate computation of expectation values. We use this strategy to study the phase diagram of a $\mathbb{Z}_{2}$ gauge theory with Higgs matter.

\begin{figure*}
\includegraphics[width=0.9\textwidth]{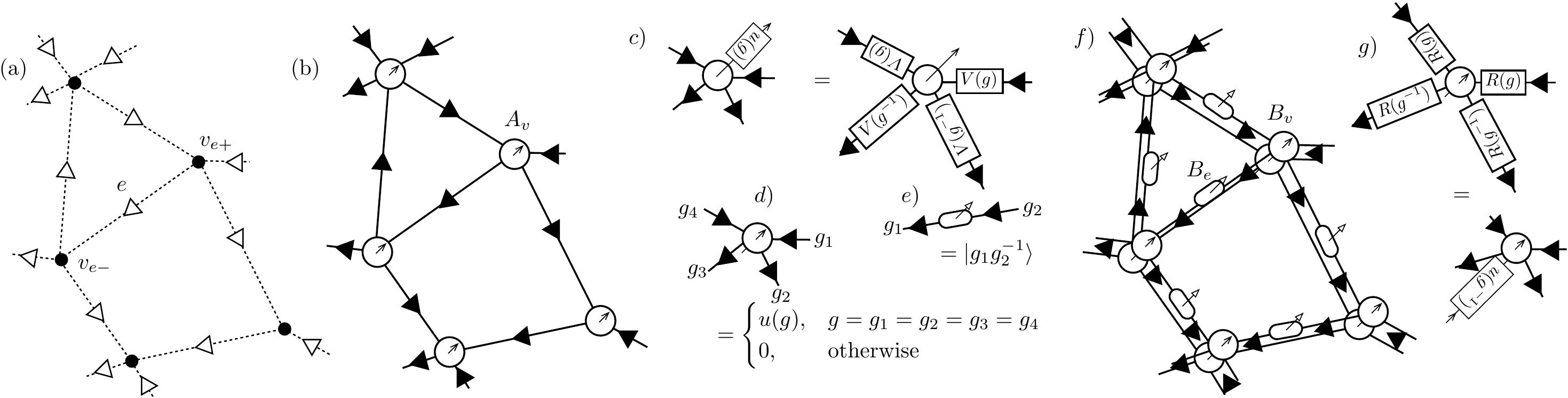}
\caption{(a) Definition of the graph $\Lambda$ with vertices $v$ and oriented edges $e$. (b) Construction of the PEPS $\ket{\psi(A)}$ from tensors $A_v$ associated to the vertices $v$ and with virtual bonds along the edges $e$ and physical indices depicted as arrows pointing out of the center of every tensor. (c) Symmetry of a PEPS tensor to ensure global symmetry of the state $\ket{\psi(A)}$ under the group action $U_{g}$. (d) Definition of the tensor $X_{v}$ used in the construction of the projector $P$ onto the gauge-invariant subspace. (e) Result of acting with the tensor $X_{e}$ on the physical input state $\ket{1}$, which is the only case we need throughout this manuscript. (f) PEPS $\ket{\Psi(B)}$ with vertex tensors $B_v$ and edge tensors $B_e$ obtained from acting with $P$ on $\ket{\psi(A)}\bigotimes_{e}\ket{1}_{e}$. (g) Symmetry property of the tensor $X_v$.}
\label{fig:defs}
\end{figure*}

Throughout this manuscript we consider a lattice or, more generally, a graph, $\Lambda$, with quantum degrees of freedom living on the vertices $v$, to which we henceforth refer as the matter fields. To every vertex $v\in\Lambda$, there is an associated Hilbert space $\hilbert_v$, such that the total quantum state of the matter fields lives in the Hilbert space $\hilbert_{\Lambda}^{(\text{m})}=\bigotimes_{v\in\Lambda} \hilbert_v$. We furthermore decorate $\Lambda$ with oriented edges $e$ as in Fig.~\ref{fig:defs}(a). With a slight abuse of notation, we denote the edges in $\Lambda$ as $e\in\Lambda$, where the only difference with the vertices $v\in\Lambda$ is in the chosen character. For every vertex $v$, we denote $E^+_v$ as the set of outgoing edges and $E^-_v$ as the set of incoming edges. Correspondingly, we define $v_{e\pm}$ as that vertex for which $e\in E_{v}^{\pm}$, such that edge $e$ points from $v_{e+}$ to $v_{e-}$.

We start from a quantum many body state $\ket{\psi}\in\hilbert^{(\text{m})}$
for the matter fields, which is invariant under the global action $U_{\Lambda}(g)=\bigotimes_{v\in \Lambda} U_v(g)$ of elements $g$ in a finite or compact symmetry group $\mathsf{G}$, \textit{i.e.} $U_{\Lambda}(g) \ket{\psi} = \ket{\psi}$. Here, $U_{v}(g)$ corresponds to a unitary representation of $\mathsf{G}$ on the local Hilbert space $\hilbert_v$ of site $v$. In order to transform this state into a new state that is invariant under a local action of $\mathsf{G}$, we introduce new degrees of freedom, the gauge fields. We thereto define on every edge $e$ of the graph $\Lambda$ a new physical Hilbert space $\hilbert_e=\mathbb{C}[\mathsf{G}]$ \footnote{For continuous groups with the structure of a differentiable manifold, i.e.\ Lie groups, the notation $L^{2}(\mathsf{G})$ is more common and correct.}, spanned by the `position' basis $\{\ket{g}\}_{g\in\mathsf{G}}$. The left and right group action of $\mathsf{G}$ on $\hilbert_e$ is given by 
\begin{align}
L_{e}(h)\ket{g}_{e}&=\ket{hg}_{e},& R_{e}(h) \ket{g}_{e}&=\ket{g h^{-1}}_{e}.
\end{align}
We denote the total gauge field Hilbert space as $\hilbert_{\Lambda}^{(\text{g})}=\bigotimes_{e\in\Lambda} \hilbert_e$ and the combined gauge matter Hilbert space as $\hilbert_{\Lambda}^{(\text{g,m})}=\hilbert_{\Lambda}^{(\text{g})}\otimes \hilbert_{\Lambda}^{(\text{m})}$. A local gauge transformation with group element $g$ on vertex $v$ corresponds to the unitary operator $U_v(g) \bigotimes_{e\in E_{v}^{+}} R_e(g) \bigotimes_{e'\in E_{v}^{-}} L_{e'}(g)$. Group averaging using the Haar measure \cite{1995JMP....36.6456A} can then be used to build a local projector $P_{v}$ onto the invariant subspace, i.e.\ the states satisfying `Gauss law` at vertex $v$, 
\begin{equation}
P_v =\int \mathrm{d} g_{v} U_v(g_v) \bigotimes_{e\in E_{v}^{+}} R_e(g_{v}) \bigotimes_{e'\in E_{v}^{-}} L_{e'}(g_{v}).
\end{equation}
Note that $[P_v,P_{v'}]=0$ thanks to $[L_e(g),R_e(h)]=0$, so that the projector onto the gauge-invariant subspace of a region $\Gamma$ is defined as $P_{\Gamma}=\prod_{v\in\Gamma} P_{v}$. In particular, $P= P_{\Lambda}$ is the projector onto the gauge-invariant subspace of $\hilbert_{\Lambda}^{(\text{g,m})}$, corresponding to the physical Hilbert space $\hilbert^{(\text{phys})}_{\Lambda}$.

With these ingredients, we now present our prescription for gauging quantum states. It is given by the linear map $G:\hilbert^{(\text{m})}_{\Lambda}\to\hilbert^{(\text{phys})}_{\Lambda}$ that acts on states $\ket{\psi}\in\hilbert^{(\text{m})}_{\Lambda}$ as $G\ket{\psi} = P \ket{\psi} \bigotimes_{e} \ket{1}_{e}$.
We thus construct the direct product of the original state $\ket{\psi}$ for the matter field with a state for the gauge field which is a product state of $\ket{1}_{e}$ on every edge $e$ corresponding to the identity element $g=1$ of the group. The result is then projected into the gauge invariant subspace $\hilbert^{(\text{phys})}_{\Lambda}$ by $P$. We can explicitly evaluate $G$ and find
\begin{equation}
G\ket{\psi} =  \prod_{v\in\Lambda} \int \mathrm{d} g_v U_v(g_v) \ket{\psi}\bigotimes_{e} \ket{ g_{v_{e-}}g_{v_{e+}}^{-1}}_{e}\label{eq:gauging}.
\end{equation}
From this definition, it is clear that $G U(g)=G$, since a global transformation $g_{v}\to g_{v} g$ will not appear in the configuration of gauge fields on the edges, if every edge is connecting two vertices $v_{e+}$ and $v_{e-}$. One can in fact check that $G^{\dagger}G=\int \mathrm{d}g\, U_\Lambda(g)$ is the projector onto the trivial representation of the global symmetry group $\mathsf{G}$ in $\hilbert^{(\text{m})}_{\Lambda}$. This implies that initial states $\ket{\psi}$ that transform under a non-trivial representation of the global symmetry $\mathsf{G}$ are annihilated by the gauging process $G$. This is the mathematical equivalent of the well known fact that one cannot have a total net charge in a gauge theory on a closed surface \footnote{The only way around is to have one or more dangling edges which are only connected to one vertex, which is known as a rough boundary \cite{2012CMaPh.313..351K}.}.

We now look for an associated operator map $\mathscr{G}$ for gauging an arbitrary matter operator $O$, in such a way that $G(O\ket{\psi})=\mathscr{G}[O] G\ket{\psi}$. Since $G^\dagger G\ket{\psi}=\ket{\psi}$, where $\ket{\psi}$ is assumed to be invariant under the global symmetry action, one could define $\mathscr{G}[O]=G O G^\dagger$. However, for an operator $O$ with non-trivial support on a compact region $\Gamma$, the resulting gauged operator $G O G^\dagger$ would have non-trivial support on the whole lattice, \textit{i.e.} it would no longer be locally supported. We therefore want to construct a different gauging map $\mathscr{G}_{\Gamma}$ which maps local matter operator to local gauge-matter operators. For any $\Gamma\subset\Lambda$ containing both vertices $v_{e\pm}$ of all of its edges, but not necessarily all edges of its vertices, we first introduce the operator map $\mathscr{P}_{\Gamma}: \mathbb{L}(\hilbert_{\Gamma}^{(\text{g,m})})\to \mathbb{L}(\hilbert_{\Gamma}^{(\text{g,m})})$ as
\begin{displaymath}
\begin{split}
\mathscr{P}_{\Gamma}[O]=\int \prod_{v\in\Gamma}\mathrm{d}g_v \big[\prod_{v\in\Gamma} U_v(g_v) \prod_{e\in\Gamma} L_e(g_{v_{e-}})R_{e}(g_{v_{e+}})\big]\\
O \big[\prod_{v\in\Gamma} U_v(g_v) \prod_{e\in\Gamma} L_e(g_{v_{e-}})R_{e}(g_{v_{e+}})\big]^{\dagger}.
\end{split}
\end{displaymath}
Note that $\mathscr{P}_{\Gamma}[O] P_{v}=P_{v} \mathscr{P}_{\Gamma}[O]$ for any $v\in\Lambda$, so that $\mathscr{P}_{\Gamma}$ produces gauge-invariant operators, even though it does not include an explicit projector onto the gauge-invariant subspace, i.e.\ it does not necessarily annihilate states which are not gauge-invariant. In particular, $\mathscr{P}_{\Gamma}[1]=1$. We can then also define the gauging map 
\begin{displaymath}
\mathscr{G}_{\Gamma}:\mathbb{L}(\hilbert_{\Gamma}^{(\text{m})})\to \mathbb{L}(\hilbert_{\Gamma}^{(\text{g,m})}):O\to \mathscr{P}_{\Gamma}[O\bigotimes_{e\in\Gamma} \ket{1}_{e}\bra{1}_{e}].
\end{displaymath}
One can check that $\mathscr{G}_{\Gamma}[O]$ acts diagonally on the gauge degrees of freedom, in such a way that
\begin{displaymath}
\mathscr{G}_{\Gamma}[O]\bigotimes_{e\in\Gamma} \ket{1}_{e} = \bigotimes_{e\in\Gamma} \ket{1}_{e} \otimes \int\mathrm{d}g U_{\Gamma}(g) O U_{\Gamma}(g)^{\dagger}.
\end{displaymath}
Combining this property with $P_v\mathscr{G}_{\Gamma}[O]=\mathscr{G}_{\Gamma}[O] P_v$ for any $v\in\Lambda$, it is easy to show that this map indeed satisfies $G O\ket{\psi}=\mathscr{G}_{\Gamma}[O] G\ket{\psi}$ for symmetric operators ($[O,U_{\Gamma}(g)]=0, \forall g\in\mathsf{G}$), where the support of the gauged operator $\mathscr{G}_{\Gamma}[O]$ is equivalent to the support of the original matter operator (but of course also contains the gauge degrees of freedom on the edges $e\in\Gamma$). In addition, this allows to easily show that $\mathscr{G}_{\Gamma}$ is invertible onto the space of symmetric operators using the expected prescription 
\begin{displaymath}
\tr_{(\text{g})}\big[\mathscr{G}_{\Gamma}[O]\bigotimes_{e\in\Gamma} \ket{1}_{e}\bra{1}_{e}\big]=\int\mathrm{d}g U_{\Gamma}(g) O U_{\Gamma}(g)^{\dagger}=O,
\end{displaymath}
where $\tr_{(\text{g})}$ is a partial trace over the gauge degrees of freedom living at the edges $e\in\Gamma$.

The current gauging procedure generates a gauged state $\ket{\Psi}=G\ket{\psi}$ at a zero value of the gauge coupling constant, i.e. the gauge degrees of freedom are frozen so that there are no magnetic fluxes and the gauged theory produces equivalent expectation values as the original theory. To introduce gauge dynamics for nonzero values of the coupling constant, we could manually add the electric energy term to the Hamiltonian. However, since we are working at the level of quantum states, we follow a different approach. Instead, we apply the well-known local filtering operation \cite{2004AnPhy.310..493A,2005AnPhy.318..316C,PhysRevB.77.054433}
\begin{equation}
\ket{\Psi}\to \prod_{e\in\Lambda} \exp(-\frac{\beta}{2} \mathcal{E}_{e}^2)\ket{\Psi}\label{eq:filter}
\end{equation}
with $\mathcal{E}_{e}$ the electric field operator on edge $e$. It is now easy to check that the `ungauging' process (set $\beta=0$ and project the gauge fields on the links $e$ in $G\ket{\psi}$ onto $\ket{1}_{e}$) results in $\int\mathrm{d}g\,U_{\Lambda}(g)\ket{\psi}=\ket{\psi}$, where the last equality only holds if the starting state was invariant under the symmetry action.

It turns out that this gauging procedure is very natural in the framework of PEPS. Let us hereto introduce the PEPS $\ket{\psi(A)}$ using tensors $A_v$ associated to every vertex $v\in\Lambda$. These tensors act as a multilinear map from virtual vector spaces $\mathbb{V}_e$ associated to the incoming edges $e\in E_{v}^{-}$ to the virtual vectors spaces $\mathbb{V}_{e'}$ associated to the outgoing edges $e'\in E_{v}^{+}$ and the physical Hilbert space $\hilbert_v$. We identify $\mathbb{V}_e$ with $\mathbb{C}^{D_e}$ with $D_{e}$ the bond dimension on edge $e$. By choosing a canonical basis in all vector spaces, we can write
\begin{equation}
A_v=\sum_{s,\{\alpha_{e'}\},\{\beta_{e}\}} (A_v)^s_{\{\alpha_{e'}\},\{\beta_{e}\}} \ket{s}\bigotimes_{e'\in E_{v}^{+}} \rket{\alpha_{e'}} \bigotimes_{e\in E_{v}^{-}} \rbra{\beta_{e}}.\label{eq:tensor}
\end{equation}
The physical state $\ket{\psi(A)}$ is obtained by contracting the corresponding kets and bras of all virtual spaces. This construction is illustrated in Fig.~\ref{fig:defs}(b). We now assume that the PEPS tensors $A_{v}$ satisfy the generic property of \emph{injectivity} \cite{2007arXiv0707.2260P}, meaning that the there exists a finite region $\Gamma\subset \Lambda$ such that the map from virtual boundary $\mathbb{V}_{\partial\Gamma}=\bigotimes_{e\in\partial\Gamma}\mathbb{V}_{e}$ to physical bulk $\hilbert^{(\text{m})}_{\Gamma}=\bigotimes_{v\in\Gamma}\hilbert_{v}$ is injective. This property guarantees that the PEPS is `well-behaved', e.g.\ that it is the unique ground state of a local parent Hamiltonian. If an injective PEPS $\ket{\psi(A)}$ is invariant under the global action $U_{\Lambda}(g)$ for $g\in\mathsf{G}$, then it was proven in Ref.~\onlinecite{1367-2630-12-2-025010} that there must exist (projective) representations $V_e$ of $\mathsf{G}$ on the virtual spaces $\mathbb{V}_{e}$ such that $A_v$ acts as an intertwiner
\begin{equation}
A_v \bigg[\bigotimes_{e'\in E_v^-} V_{e'}(g)\bigg]= U_v(g) \bigg[\bigotimes_{e\in E_{v}^{+}} V_{e}(g)\bigg] A_v.\label{eq:intertwiner}
\end{equation}
A slightly different form of this equation is presented in Fig.~\ref{fig:defs}(c). While the representations $V_e$ are not required to be unitary, we can in principle perform a `gauge' transformation \footnote{This transformation lives completely at the virtual level and its existence is a mere consequence of the PEPS representation. It bears no relation to the physical gauge fields, which at this point still have to be introduced.} on the PEPS tensors to transform any finite-dimensional representation to a unitary representation if $\mathsf{G}$ is a compact group.

The projector onto the gauge invariant subspace of $\hilbert_{\Lambda}^{(\text{g,m})}$ also has a simple tensor network description by introducing virtual spaces $\mathbb{V}_e'\equiv \mathbb{C}[\mathsf{G}]$ on every edge $e$ and contracting all virtual bonds of vertex tensors $X_v:\hilbert_v \bigotimes_{e\in E_{v}^{-}} \mathbb{V}_{e}' \to \hilbert_v \bigotimes_{e\in E_{v}^{+}} \mathbb{V}_e'$ given by
\begin{equation}
X_v= \int \mathrm{d} g \ U_v(g)\bigotimes_{e'\in E_{v}^{+}} \rket{g}_{e'} \bigotimes_{e\in E_{v}^{-}} \rbra{g}_{e},
\end{equation}
as sketched in Fig~\ref{fig:defs}(d), and edge tensors $X_e: \hilbert_e\otimes \mathbb{V}_{e}' \to \hilbert_{e}\otimes \mathbb{V}_{e}'$ given by 
\begin{equation}
X_e= \int \mathrm{d} g_{-}\mathrm{d} g_{+} \ L_e(g_{-}) R_e(g_{+}) \otimes \rket{g_-}\rbra{g_{+}}.
\end{equation}
For the case of continuous groups, the virtual dimensions of this tensor network are infinite and this representation is not amenable to numerical computations. Similar constructions of $P$ appeared in the context of spin networks \cite{1994gr.qc....11007B}, and recently in the context of tensor networks \cite{2014arXiv1404.7439S,2014arXiv1405.4811T}, where it was also discussed how to compress the bond dimension to finite values.

Applying this gauging procedure $G$ to the symmetric PEPS $\ket{\psi(A)}\in\hilbert^{(\text{m})}_{\Lambda}$, whose tensors $A$ satisfy Eq.~\eqref{eq:intertwiner}, results in a new PEPS with virtual spaces given by $\mathbb{W}_{e}=\mathbb{V}_{e}\otimes\mathbb{V}'_{e}$. Indeed, we can write the new state $G\ket{\psi(A)}$ as a PEPS $\ket{\Psi(B)}$ sketched in Fig.~\ref{fig:defs}(g) with vertex tensors
\begin{equation}
B_v= \int\mathrm{d}g \ U_v(g) A_v \bigotimes_{e'\in E_{v}^{+}} \rket{g}_{e'}\bigotimes_{e\in E_{v}^{-}}\rbra{g}_{e}
\end{equation}
and edge tensors given by
\begin{equation}
B_e=\int\mathrm{d} g_{+}\mathrm{d} g_{-} \ket{g_{-} g_{+}^{-1}} \otimes \openone_{D_e} \otimes \rket{g_{-}}_e \rbra{g_{+}}_e
\end{equation}
where the first ket corresponds to the physical state, the second factor to the action on $\mathbb{V}_{e}$ and the last factor to the action on the $\mathbb{V}'_{e}$. Using the intertwining property of $A_v$ in Eq.~\eqref{eq:intertwiner} or Fig.~\ref{fig:defs}(c) and the symmetry property of $X_{v}$ sketched in Fig.~\ref{fig:defs}(g), one can check that
\begin{displaymath}
[\bigotimes_{e'\in E_{v}^{+}} V_{e'}(g^{-1})\otimes R'_{e'}(g^{-1})] B_v [\bigotimes_{e\in E_{v}^{-}} V_e(g)\otimes R'_e(g)] = B_v,
\end{displaymath}
where all factors act on the virtual level $\mathbb{W}_{e}=\mathbb{V}_{e}\otimes \mathbb{V}'_{e}$. In particular, $R'_{e}(g)$ corresponds to the right group action of $\mathsf{G}$ on the virtual space $\mathbb{V}'_{e}$. We similarly have that 
\begin{displaymath}
[V_e(g^{-1})\otimes R'_e(g^{-1})] B_e [V_e(g) \otimes R'_e(g)]=B_e.
\end{displaymath}
This implies that the resulting PEPS cannot be injective, but below we prove for it to be $\mathsf{G}$-injective \cite{2010AnPhy.325.2153S} instead. This property means that the map from virtual boundary to physical bulk is only invertible up to the action of group $\mathsf{G}$ ---whose representation on $\mathbb{W}_{e}=\mathbb{V}_{e}\otimes\mathbb{V}_{e}'$ is here given by $V_{e} \otimes R'_{e}$--- and is intricately related to topological order. More specifically, for a discrete group $\mathsf{G}$, the property of $\mathsf{G}$-injectivity allows for the presence of anyonic excitations, although they could of course be confined or condensed depending on the matter interactions and the gauge coupling constant. We refer to the Supplementary Material for additional details.

For the proof, we consider a region $\Gamma$ on which the PEPS tensors act as an injective map. The range of this map is denoted as $\mathbb{A}\subset\hilbert_{\Gamma}^{(\text{m})}$ with $\dim\mathbb{A}=\dim\mathbb{V}_{\partial\Gamma}$ and corresponds to the support of the reduced density matrix of $\ket{\psi(A)}$ in $\Gamma$. Let $\{\ket{\phi_{i}},i=1,\ldots,\text{dim}\mathbb{A}\}$ be an orthonormal basis for this subspace, where every $\ket{\phi_{i}}$ is obtained from a unique state $\rket{\tilde{\phi}_{i}}\in \mathbb{V}_{\partial\Gamma}$ on the virtual boundary. A frustration free parent Hamiltonian can be constructed from terms $h_{\Gamma}^{(\text{m})}=1-\sum_{i} \ket{\phi_{i}}\bra{\phi_{i}}$, \textit{i.e.} the projector onto the orthogonal complement of $\mathbb{A}$. The symmetry under $\mathsf{G}$ follows from the fact that $U_{\Gamma}(g)\ket{\phi_{i}}=\sum_{j} u_{j,i}(g) \ket{\phi_j}$ with $u(g)$ a unitary representation whose matrix elements are given by $u_{j,i}(g)=\rbraket{\tilde{\phi}_{j}|V_{\partial \Gamma}(g)|\tilde{\phi}_{i}}$, where $V_{\partial\Gamma}$ is the tensor product representation of the different $V_{e}$ representations on the virtual boundary $\mathbb{V}_{\partial \Gamma}$.

For the gauged PEPS with tensors $B$, we choose $\Gamma$ such that it excludes the physical spaces of the gauge fields on the edges $e\in\partial \Gamma$. We denote by $\Gamma^{\circ}$ the set of vertices in the interior of $\Gamma$, \textit{i.e.} those vertices for which all edges and neighbouring vertices are also contained in $\Gamma$. The vertices in the set $\Delta \Gamma=\Gamma \setminus \Gamma^{\circ}$ are on the (inside) boundary and have one edge $e\in\partial \Gamma$ \footnote{For simplicity, we ignore the corner case where a vertex $v\in\Delta \Gamma$ has more than one edge $e\in\partial \Gamma$, as this brings along additional complications with the injectivity construction. This corner case can be avoided on the hexagonal lattice but not on the square lattice.}. If we now define the state $\rket{\tilde{\Phi}_{i,\{g_{v}\}}}=\rket{\tilde{\phi}_{i}}\bigotimes\rket{g_{v_e}}_{e}$ on the boundary $\mathbb{W}_{\partial\Gamma}=\bigotimes_{e\in\partial\Gamma} \mathbb{V}_{e} \otimes \mathbb{V}'_{e}$, where every edge $e$ has a one to one correspondence with a vertex $v_{e}\in\Delta \Gamma$, the resulting state in the bulk $\hilbert_{\Gamma}^{(\text{g,m})}$ is given by
\begin{displaymath}
\begin{split}
\ket{\Phi_{i,\{g_{v}\}}}=\prod_{v\in\Delta \Gamma} U_{v}(g_v) \bigotimes_{e'\in E_{v}^{+}\cap \Gamma} R_{e}(g_{v}) \bigotimes_{e\in E_{v}^{-}\cap \Gamma} L_{e}(g_{v})\\
\prod_{v'\in\Gamma^{\circ}} P_{v'}\ket{\phi_{i}}\bigotimes_{e\in\Gamma} \ket{1}_{e}.
\end{split}
\end{displaymath}
One can show that $\braket{\Phi_{i,\{g_v\}}|\Phi_{i',\{g_{v}'\}}}=0$ if there is no $g\in\mathsf{G}$ such that $\{g_{v}'\}=\{g_{v} g\}$. The reason for this is that the edges along the boundary, between two vertices in $\Delta\Gamma$, allow to resolve the elements $g_{v}$ up to the global transformation $g_v\to g_vg$. Having resolved $g_{v}$ up to a factor $g$, the edges connecting $\Delta \Gamma$ to $\Gamma^{\circ}$ will act as a rough boundary, so that the inner product of the edge degrees of freedom will force all interior gauge transformations in ket and bra to be equal up to the global transformation $g$, resulting in 
\begin{multline*}
\braket{\Phi_{i,\{g_v\}}|\Phi_{i',\{g_{v}g\}}}=\braket{\phi_{i}|U_{\Gamma}(g)|\phi_{i'}}\\
=\rbraket{\tilde{\phi}_{i}|V_{\partial \Gamma}(g)|\tilde{\phi}_{i'}}=u_{i,i'}(g).
\end{multline*}
Hence, the preimage of every bulk state $\ket{\Phi_{i,\{g_{v}\}}}$ is the set of states
\begin{displaymath}
\{\bigotimes_{e\in\partial\Gamma} V_{e}(g)\otimes R_{e}'(g)\rket{\tilde{\Phi}_{i,\{g_{v}\}}},\forall g\in\mathsf{G}\},
\end{displaymath}
in line with the concept of $\mathsf{G}$-injectivity \cite{2010AnPhy.325.2153S}. Even though the set of $\{\ket{\Phi_{i,\{g_{v}\}}},\forall i,\forall g_v \in \mathsf{G}, \forall v\in \Delta \Gamma\}$ is overcomplete, we can still check that
\begin{displaymath}
h_{\Gamma}^{(\text{g,m})}=1-\sum_{i}\int\prod_{v\in\Delta \Gamma} \mathrm{d} g_{v}\,\ket{\Phi_{i,\{g_v\}}}\bra{\Phi_{i,\{g_{v}\}}}
\end{displaymath}
is a projector that annihilates the PEPS $\ket{\Psi(B)}$. Lets now try to rewrite this parent Hamiltonian using the operator gauging map $\mathscr{G}_{\Gamma}$. Note that
\begin{displaymath}
\sum_{i}\int\prod_{v\in\Delta \Gamma} \mathrm{d} g_{v}\,\ket{\Phi_{i,\{g_v\}}}\bra{\Phi_{i,\{g_{v}\}}}
\neq \mathscr{G}_{\Gamma}[\sum_{i} \ket{\phi_{i}}\bra{\phi_{i}}]
\end{displaymath}
since the left hand side contains two independent integrations for every interior vertex $v\in\Gamma^{\circ}$. Instead, we find $1-h_{\Gamma}^{(\text{g,m})}=\mathscr{G}_{\Gamma}[1-h_{\Gamma}^{(\text{m})}]\prod_{v\in\Gamma^{\circ}} P_{v}$ so that $h_{\Gamma}^{(\text{g,m})}$ contains an explicit energy penalty for all non gauge-invariant states. Since the physical Hilbert space $\hilbert^{(\text{phys})}_\Gamma$ of a gauge theory is restricted to gauge invariant states satisfying the `Gauss law` constraint on every vertex, we can safely omit this additional factor and instead write
\begin{equation}
h_{\Gamma}^{(\text{g,m})}=\mathscr{G}_{\Gamma}\bigg[h_{\Gamma}^{(\text{m})}\bigg]+
\left(1-\mathscr{P}_{\Gamma}\bigg[\bigotimes_{e\in\Gamma}\ket{1}_{e}\bra{1}_{e}\bigg]\right)\label{eq:gaugedhamiltonian}
\end{equation}
We can recognise the first term as the gauged matter Hamiltonian, whereas the second term is a pure gauge term. One can verify that it acts as a projector giving an energy penalty $1$ to states with nonzero magnetic flux through any plaquette contained in $\Gamma$, which has to contain at least a single plaquette for the injectivity construction. Hence, for a single plaquette $p$, we can then write the second term as
\begin{equation}
1-\int \prod_{e\in\partial p} \mathrm{d}g_{e} \ket{g_{e}}_{e}\bra{g_{e}}_{e} \chi^{(\text{reg})}\big(\prod_{e} g_{e}\big) \label{eq:plaquette}
\end{equation}
with $\chi^{(\text{reg})}(g)=\delta(g-1)$ the character of the regular representation, where the product in its argument is ordered in the way the edges $e$ appear along the boundary $\partial p$ of the plaquette $p$, and all edges are assumed to be oriented similarly. This term corresponds exactly to the magnetic term of the quantum double models \cite{2003AnPhy.303....2K}. The typical magnetic term from the Kogut-Susskind lattice gauge Hamiltonian 
would replace $\chi^{(\text{reg})}$ with $\mathrm{Re}\ \chi^{(l)}$, with $l$ the fundamental representation in the case of a Lie group $\mathsf{G}$, but has the same ground state subspace. Indeed, one can check that throughout our gauging construction, by initialising the gauge fields in the $\ket{1}$ configuration, we are effectively working at zero coupling constant for the gauge field and the magnetic energy term is automatically minimised.

Finally, the filtering operation in Eq.~\eqref{eq:filter} can be applied to the PEPS without increasing the bond dimension or changing the $\mathsf{G}$-injectivity property. If $H^{(\text{g,m})}$ is a frustration free Hamiltonian with terms $h^{(\text{g,m})}$ that annihilate the ground state, then the filtered PEPS is the ground state of a parent Hamiltonian built of terms
\begin{displaymath}
e^{\frac{\beta}{2} \sum_{e\in\Gamma}\mathcal{E}_{e}^2} h_{\Gamma}^{(\text{g,m})} e^{\frac{\beta}{2}\sum_{e\in\Gamma} \mathcal{E}_{e}^2}=h_{\Gamma}^{(\text{g,m})}+\beta\sum_{e\in\Gamma} \mathcal{E}_{e}^2+\ldots
\end{displaymath}
where the terms in $\ldots$ can be expected to become irrelevant under renormalization for small $\beta$, as they correspond to higher-dimensional operators in the continuum theory.

We now apply this gauging procedure for quantum states and operators to a number of examples. Consider as a first consistency check a nearest neighbor pair of vertices $\Gamma=\{v_{-},v_{+}\}$ with corresponding edge $e=(v_{-},v_{+})$. Let $O^{i}$ be a vector of operators such that $U(g) O^{i} U(g)^{\dagger}= \phi^{j,i}(g) O^{j}$ with $\phi$ some unitary representation of $\mathsf{G}$. Consider $O=\sum_{i} O^{i}_{v_{-}} O^{i\dagger}_{v_{+}}$. We obtain $\mathscr{G}_{\Gamma}[O]=\sum_{j,k} O^{j}_{v_{-}} \Phi^{j,k}_e O^{k\dagger}_{v_{+}}$ where $\Phi^{j,k}$ is given by
\begin{displaymath}
\Phi^{j,k}=\int \mathrm{d} g\, \phi^{j,k}(g)\ket{g}\bra{g},
\end{displaymath}
\text{i.e.} it is the operator that extracts the $(j,k)$ element of the representation $\phi$. We thus recover the `minimal coupling' rule for \textit{e.g.} a hopping term. This example trivially generalises to the case where $\Gamma$ contains a path between two distant vertices, as would be the case for a correlation function. The map $\mathscr{G}_{\Gamma}$ then creates a gauge-invariant correlation function by inserting a Wilson line along the path. The choice of path will be irrelevant as long as the state is an exact ground state of the plaquette operators, \textit{i.e.}\ as long as there are no fluxes created by \textit{e.g.}\ a finite value of the gauge coupling constant.  For more complex matter Hamiltonians with for example plaquette interactions, one can check that our prescription exactly reproduces the gauging construction used in Ref.~\onlinecite{PhysRevB.86.115109} to establish the relation between symmetry protected topological order \cite{PhysRevB.87.155114} and the twisted quantum double models \cite{PhysRevB.87.125114}. In Ref.~\onlinecite{2014arXiv1412.5604W} the corresponding effect is investigated at the level of the PEPS which, after applying the gauging prescription here developed, acquires the property of twisted $\mathsf{G}$-injectivity \cite{2013arXiv1307.7763B} or, more generally, MPO-injectivity \cite{2014arXiv1409.2150B}.

\begin{figure}
\includegraphics[width=\columnwidth]{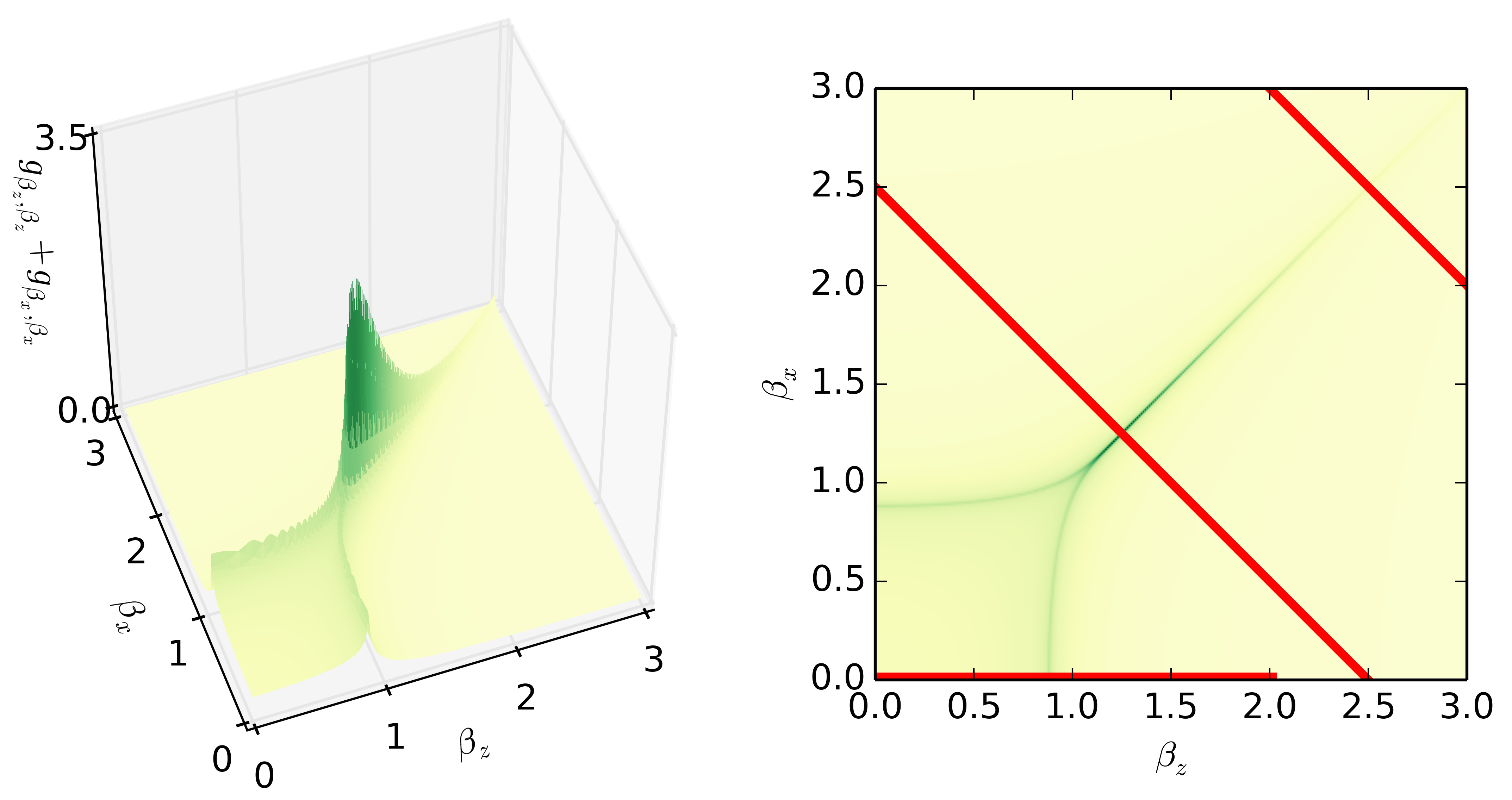}
\caption{Trace of the metric $g(\bm{\beta})$ with $\bm{\beta}=(\beta_z,\beta_x)$, as obtained from the fidelity $\lvert \braket{\Psi_{\bm{\beta}}|\Psi_{\bm{\beta}+\bm{\delta\beta}}}\rvert=\exp(-N \bm{\delta \beta}^{\mathrm{T}} g(\bm{\beta}) \bm{\delta\beta})$ where $N$ is the (infinite) number of sites, as defined in Ref.~\onlinecite{PhysRevLett.99.100603}. An analytic expression for $g$ along the coordinate axes ($\beta_z=0$ or $\beta_x=0$) was obtained in Ref.~\onlinecite{PhysRevA.78.010301}. The red lines on the right panel indicate the slices studied in Fig.~\ref{fig:fid} and Fig.~\ref{fig:expval}.}
\label{fig:dfid}
\end{figure}
As a more elaborate example, we now consider the phase diagram of a gauge theory with Higgs matter, i.e.\ scalar bosonic matter transforming non-trivially under the gauge group, for the specific case of $\mathsf{G}=\mathbb{Z}_2=\{1,-1\}$ in $(2+1)$ dimensions. Using a basis $\ket{1},\ket{-1}$ for both $\hilbert_{v}$ and $\hilbert_{e}$ we have $L_e(-1)=R_e(-1)=\tau^x$ and we also choose to have $U_v(-1)=\sigma^x$, with $\sigma^x,\tau^x$ the Pauli operators for matter and gauge fields respectively. We start from the ground state of the Ising model 
\begin{displaymath}
H_{I}=\sum_{v}(1-\sigma^x_{v})+\beta_{z} \sum_{e} (1-\sigma^z_{v_{e-}}\sigma^z_{v_{e+}})
\end{displaymath}
at $\beta_z=0$, \textit{i.e.} the state $\ket{\psi_{0}}=\prod_{v} \ket{+}_{v}$. Instead of turning on a finite $\beta_{z}$ in the Hamiltonian, we again resort to applying a filtering operation $\prod_{e}\ec^{\frac{\beta_{z}}{4}\sigma^z_{v_{e-}}\sigma^z_{v_{e+}}}$. The resulting state $\ket{\psi_{\beta_z}}$ is a PEPS with bond dimension $2$ and parent Hamiltonian \cite{PhysRevLett.96.220601}
\begin{displaymath}
H^{(\text{m})}=\sum_{v} (\ec^{-\frac{\beta_{z}}{2}\sum_{e\in E_{v}}\sigma^z_{v_{e+}}\sigma^z_{v_{e-}}}-\sigma^x_v),
\end{displaymath}
where $E_v=E_v^{+}\cup E_{v}^{-}$. This matches the Ising Hamiltonian $H_{I}$ at lowest order in $\beta_{z}$. We now apply the gauging procedure $\ket{\Psi_{\beta_{z},0}}=G\ket{\psi_{\beta_{z}}}$ and switch on an electric field $\mathscr{E}^{2}=(1-\tau^x)/2$ with coupling constant $\beta_x$ using local filtering, to obtain the state
\begin{displaymath}
\ket{\Psi_{\beta_{z},\beta_{x}}}=\prod_{e} \ec^{\frac{\beta_{x}}{4}\tau^{x}_{e}} \prod_{v} P_{v} \prod_{e}\ket{1}_{e}\ec^{\frac{\beta_{z}}{4}\sigma^z_{v_{e-}}\sigma^z_{v_{e+}}}\prod_{v} \ket{+}_{v}
\end{displaymath}
with $P_{v}=(1+\sigma^{x}_{v}\prod_{e\in E_{v}} \tau^{x}_{e})/2$. This state can be written as a PEPS with bond dimension $4$. However, we can easily `disentangle' the matter fields by applying a CNOT gate with the matter field as control and the gauge field as target for every pair of $(v,e\in E_{v})$. Since all these gates commute, this is a finite depth quantum circuit which transforms the Gauss law $(P_{v}-1)\ket{\Psi}=0$ into $(\tilde{P}_{v}-1)\ket{\tilde{\Psi}}=0$, with $\tilde{P}_{v}=(1+\sigma^{x}_{v})/2$ and $\ket{\tilde{\Psi}}$ the transformed state. Hence, gauge invariance in this transformed frame requires all matter fields to be in the $\ket{+}_{v}$ state so that we are left with unconstrained degrees of freedom on the edges. Applying the CNOT transformation to $\ket{\Psi_{\beta_z,\beta_x}}$ results in the state
\begin{displaymath}
\ket{\tilde{\Psi}_{\beta_{z},\beta_{x}}}=\prod_{e} \ec^{\frac{\beta_{x}}{4} \tau^{x}_{e}}\ec^{\frac{\beta_{z}}{4}\tau^{z}_{e}}\ket{\Psi_{\mathrm{TC}}} \bigotimes_{v} \ket{+}_{v} 
\end{displaymath}
with $\ket{\Psi_{\mathrm{TC}}}$ the toric code ground state \cite{2003AnPhy.303....2K} for the edge degrees of freedom. This is equivalent to the well-known correspondence between the normal $\mathbb{Z}_{2}$ gauge theory with matter, whose phase diagram was first considered by Fradkin and Shenker \cite{PhysRevD.19.3682}, and the toric code Hamiltonian with magnetic fields 
\begin{equation}
H_{h_z,h_x}=H_{\text{TC}}  -h_z\sum_{e} \tau^{z} - h_x \sum_{e} \tau^{x}_e
\end{equation}
as studied in Refs.~\onlinecite{PhysRevB.79.033109,PhysRevB.82.085114,PhysRevB.85.195104}. Note that the definition of the state $\ket{\tilde{\Psi}_{\beta_z,\beta_x}}$ depends on the order of applying the filtering in $\tau^{z}$ and in $\tau^{x}$. Since the motivation for these filtering operations comes from the lowest order in $\beta$, at which level they do commute, we can also opt for a more symmetric definition
\begin{displaymath}
\ket{\Psi'_{\beta_{z},\beta_{x}}}=\prod_{e} \ec^{\frac{\beta_{x}\tau^{x}_{e}+\beta_{z} \tau^{z}_{e}}{4}}\ket{\Psi_{\mathrm{TC}}} \bigotimes_{v} \ket{+}_{v} 
\end{displaymath}
\begin{figure*}
\includegraphics[width=0.9\textwidth]{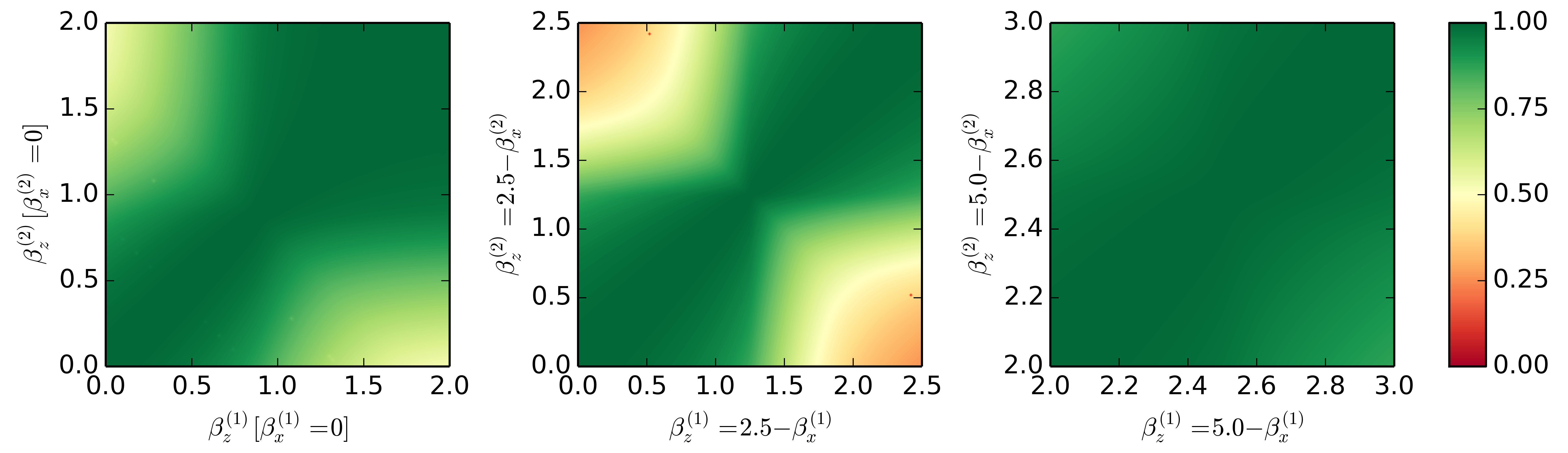}
\caption{Fidelity (per site) $\lvert\braket{\Psi_{\bm{\beta}^{(1)}}|\Psi_{\bm{\beta}^{(2)}}}\rvert^{1/N}$ between any two ground states with parameters $\bm{\beta}$ along the three different slices indicated in Fig.~\ref{fig:dfid}.}
\label{fig:fid}
\end{figure*}
\begin{figure*}
\includegraphics[width=0.9\textwidth]{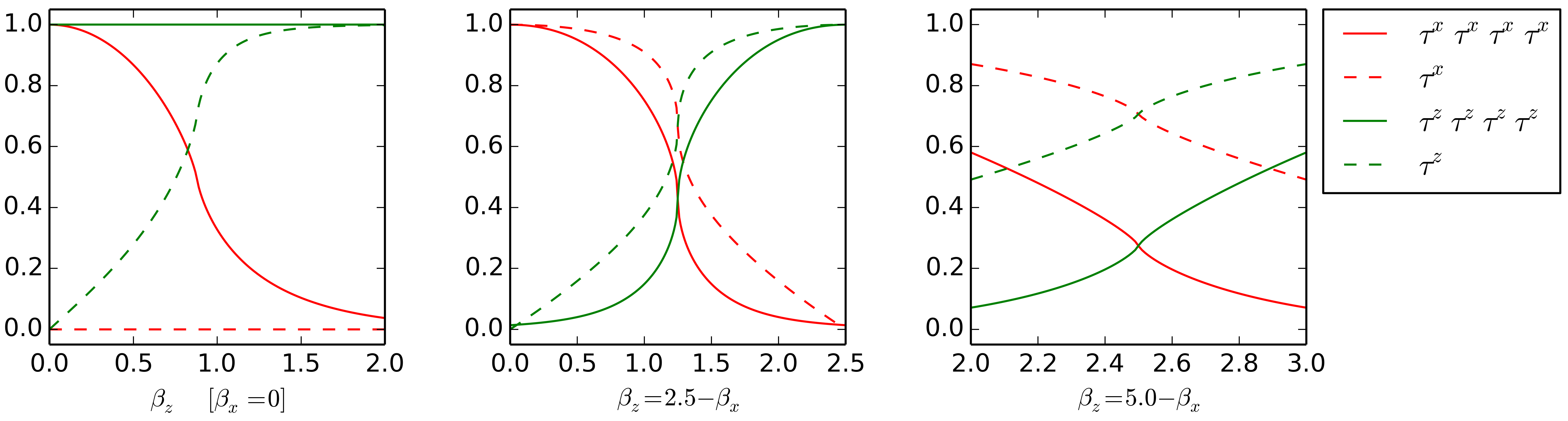}
\caption{Expectation values of the relevant Hamiltonian terms, i.e.\ $\tau^{x}$ and $\tau^{z}$ on the edges, $\tau^{x}\tau^{x}\tau^{x}\tau^{x}$ around a vertex and $\tau^{z}\tau^{z}\tau^{z}\tau^{z}$ around a plaquette, along the three different slices indicated in Fig.~\ref{fig:dfid}.}
\label{fig:expval}
\end{figure*}
The PEPS representation of $\ket{\Psi_{\text{TC}}}$ has bond dimension $2$ \cite{PhysRevLett.96.220601}, which is not increased by the local filtering. We can probe the phase diagram on this $\mathbb{Z}_{2}$ gauge theory as function of $\vec{\bm{\beta}}=(\beta_{z},\beta_{x})$ by studying the ground state fidelities in Fig.~\ref{fig:dfid} and Fig.~\ref{fig:fid}. Fidelities were computed as described in Ref.~\onlinecite{PhysRevLett.100.080601}. The fidelities illustrate that our ansatz qualitatively reproduces the phase diagram of $H_{h_x,h_z}$ and describes the same gapped phases in weak- and strong coupling limits. There is the deconfined phase with topological order around $\bm{\beta}=\bm{0}$, and the trivial Higgs phase and confined phase which are connected by a local unitary spin rotation at $\lVert\bm{\beta}\rVert\to\infty$, where $\ket{\Psi_{\beta_z,\beta_x}}$ is just a product state of eigenvectors of $\beta_z \tau^z+\beta_x\tau^x$ corresponding to the largest eigenvalue. The critical behaviour, however, is not exactly reproduced. It is known, for example, that the topological phase transition from the deconfined to the trivial phase along the coordinate axis $\beta_{z}=0$ (and by $\tau^{x}\leftrightarrow \tau^{z}$ duality also along the coordinate axis $\beta_{x}$) is in the 2D Ising universality class \cite{PhysRevB.76.224421,PhysRevB.77.054433}, whereas the corresponding phase transition of $H_{h_x,h_z}$ is in the 3D Ising universality class. Also, we obtain a second order phase transition along the duality line $\beta_{x}=\beta_{z}$ which does not seem to vanish until $\lVert \bm{\beta}\rVert=\infty$, even though the singularities in the fidelity (Figs.~\ref{fig:dfid},\ref{fig:fid}) and expectation values (Fig.~\ref{fig:expval}) become very quickly small. The corresponding transition of $H_{h_x,h_z}$ is conjectured to be first order with a discontinuity in the expectation values of the star and plaquette operator, and to end at a finite value of $\lVert \bm{\beta}\rVert$ \cite{PhysRevB.79.033109,PhysRevB.82.085114,PhysRevB.85.195104}. In contrast, the state $\ket{\Psi_{\beta_z,\beta_x}}$ seems to have continuous expectation values but a diverging derivative along a slice passing through the phase transition at the duality line $\beta_x=\beta_z$ (Fig.~\ref{fig:expval}). Note that one could also use the state $\ket{\Psi_{\beta_z,\beta_x}}$ or a slightly more general class as variational ansatz for the actual Hamiltonian, as  was done for the case $\beta_z=0$ in Ref.~\onlinecite{PhysRevB.78.205116} and quickly led to accurate results. See also Ref.~\onlinecite{Horn1980397}, where the state $\ket{\Psi_{\beta_z,\beta_x}}$ at the limiting values $\lVert\bm{\beta}\rVert=0$ and $\lVert\bm{\beta}\rVert=\infty$ was used as variational ansatz, as well as other Hamiltonian studies using large coupling expansions \cite{PhysRevD.23.2962,0305-4470-20-9-043}. A more detailed analysis of our results for $\mathbb{Z}_{2}$ and higher $\mathbb{Z}_{N}$ groups is presented elsewhere \footnote{J.~Haegeman \textit{et al}, in preparation.}.

In conclusion, we have proposed an operational procedure for gauging global symmetries at the level of individual quantum states. This procedure combines naturally with the framework of PEPS. When gauging an injective PEPS, we have shown to obtain a $\mathsf{G}$-injective PEPS and we have derived a parent hamiltonian, which automatically contains a projector version of the Kogut-Susskind hamiltonian for lattice gauge theory at zero coupling constant. By introducing gauge dynamics for nonzero values of the coupling constant using a local filtering operation, this construction results in low-parameter family of PEPS for which the phase diagram can accurately be probed, as we have illustrated by studying the phase diagram of a $\mathbb{Z}_2$ gauge theory with Higgs matter. Tensor networks are indeed promising candidates to avoid Feynman's objections \cite{Feynman:1987aa} regarding non-Gaussian states that allow to efficiently compute expectation values \cite{2010PhRvL.105y1601H}. Similar strategies are perfectly feasible for studying gauge theories with fermionic matter, using the framework of fermionic PEPS \cite{2009PhRvB..80p5129C,2010PhRvA..81a0303C,2010PhRvB..81p5104C,2010PhRvA..81e2338K,2014arXiv1404.5268P}, or even gauge theories (with or without matter) in three spatial dimensions, since no variational optimization is required \footnote{In preparation.}.

\acknowledgements{This work was initiated during the program on `Quantum Hamiltonian Complexity' held at the Simons Institute for the Theory of Computing. We acknowledge fruitful discussions with Bela Bauer, Luca Tagliacozzo, Ashley Milsted, Tobias J. Osborne, David Dudal and Henri Verschelde. Work supported by the Alexander von Humboldt
foundation (NS), the EU grants QUERG, SIQS, and by the Austrian FWF SFB grants FoQus and ViCoM. During the final stage of this work, Ref.~\onlinecite{2014arXiv1405.4811T} appeared which considers the related question of constructing gauge-invariant tensor network states for pure gauge theories.}

\clearpage

\section*{Supplementary material for ``Gauging quantum states''}

In this supplementary material, we summarize the notion of injectivity in Projected Entangled-Pair States (PEPS). Consider, thereto, a PEPS $\ket{\psi}$ which is obtained by contracting a set of local tensors $A_v$ on the vertices $v$ of a graph $\Lambda$ along matching virtual indices, which can be identified with the edges $e$ of the graph. Note that in this description, we introduce a vertex for every tensor and thus for every physical degree of freedom. In the case of a gauge theory with degrees of freedom on the edges of the physical lattice, we thus split every edge into two and introduce a new vertex in the middle.

If we consider a physical region $\Gamma$ and cut open the PEPS tensor network along the edges on $\partial \Gamma$, the tensors within the region $\Gamma$ can be `blocked' to a single tensor $A_{\Gamma}$ that maps from the virtual space $\mathbb{V}_{\partial \Gamma}=\bigotimes_{e\in\partial \Gamma} \mathbb{V}_{e}$ to the physical space $\mathbb{H}_{\Gamma}=\bigotimes_{v\in\Gamma} \mathbb{H}_{v}$, where $\mathbb{V}_{e}$ is the virtual space associated to edge $e$ and $\mathbb{H}_{v}$ is the local physical space associated to the degrees of freedom living on vertex $v$. As for typical graphs, the number of edges along $\partial\Gamma$ will grow much more slowly than the number of vertices in $\Gamma$, the map $A_{\Gamma}$ will generically become injective for sufficiently large $\Gamma$. Representing $A_{\Gamma}:\mathbb{V}_{\partial \Gamma}\to\mathbb{H}_{\Gamma}$ as a matrix, this is equivalent to the existence of a left inverse (pseudo-inverse) $A_{\Gamma}^{(-1)}$ such that $A_{\Gamma}^{(-1)}A_{\Gamma} = \openone$. If we can indeed partition the graph $\Lambda$ into a number of \emph{compact} sets $\Gamma$ (such that their size does not grow with the size of the graph) in such a way that every $A_{\Gamma}$ is injective, the resulting PEPS is called \emph{injective}. It can then be shown that this property of injectivity is conserved under further blocking, and that the resulting state is the unique ground state of a local parent Hamiltonian on every finite graph $\Lambda$ \cite{2007arXiv0707.2260P}. This parent Hamiltonian can however become gapless or have a degenerate ground space in the thermodynamic limit. The generic case of \emph{injective PEPS} corresponds to physical systems in the trivial phase, as it does not allow for e.g. a topology-dependent degeneracy of the ground state.

It was established that states within the class of the quantum double models \cite{2003AnPhy.303....2K} have a PEPS representation where the tensors $A_{\Gamma}:\mathbb{V}_{\partial \Gamma}\to\mathbb{H}_{\Gamma}$ have a purely virtual symmetry $A_{\Gamma} V_{\partial\Gamma}(g) = A_{\Gamma}$ for all $g$ in a discrete group $\mathsf{G}$. Here $V_{\partial\Gamma}(g)=\otimes_{e\in\partial\Gamma} V_{e}(g)$ and $V_{e}(g)$ is a local (projective) representation of $\mathsf{G}$ on the virtual space $\mathbb{V}_{e}$. This symmetry indicates that the PEPS cannot be injective, and a new notion of injectivity is introduced. A PEPS is called \emph{$\mathsf{G}$-injective} if there is a partition of $\Lambda$ such that every $A_{\Gamma}$ has a `pseudo-inverse' such that $A_{\Gamma}^{(-1)}A_{\Gamma} = \int\mathrm{d}g\,V_{\partial \Gamma}(g)$, \textit{i.e.}\ the projector onto the symmetric or invariant subspace under the virtual group action $V_{\partial \Gamma}(g)$. It can then be shown that this property is stable under blocking and the corresponding PEPS is the ground state of a parent Hamiltonian, whose ground state subspace on the torus has a degeneracy at the virtual level (known as a \emph{closure} of the tensor network) corresponding to the quantum double $D(\mathsf{G})$ \cite{2010AnPhy.325.2153S}. However, away from the renormalization group fixed point, the different virtual closures do not necessarily correspond to orthogonal physical states or even linearly independent physical states. In the latter case, there is no physical topological degeneracy and the system has undergone a phase transition to the trivial phase. This notion of injectivity was recently generalized to \emph{twisted $\mathsf{G}$-injectivity} \cite{2013arXiv1307.7763B} and \emph{MPO injectivity} \cite{2014arXiv1409.2150B} to also characterize those states which have a more general type of topological order. In these generalizations, the virtual symmetry $V_{\partial \Gamma}(g)$ is replaced by a matrix product operator (MPO) along the boundary $\partial \Gamma$.

In conclusion, the benefit of the PEPS description is that it enables us to study states away from exactly solvable renormalization group fixed points. In particular, the notions of injectivity here discussed allow to characterize the global properties of the resulting quantum states in terms of the virtual symmetries of the individual tensors from which the state is built.

\end{document}